\documentclass[journal]{IEEEtran}
\IEEEoverridecommandlockouts
\usepackage{cite}
\usepackage{amsmath,amssymb,amsfonts}
\usepackage[ruled,linesnumbered,lined]{algorithm2e}
\usepackage[left=0.6in, right=0.6in, bottom=0.7in, top=0.7in]{geometry}
\usepackage{graphicx}
\usepackage{textcomp}
\usepackage{xcolor}
\usepackage{bm}
\usepackage{stfloats}
\usepackage{url}
\usepackage{verbatim}
\usepackage[caption=false,font=normalsize,labelfont=sf,textfont=sf]{subfig}
\begin{document}
\title{Rate-Splitting Multiple Access for Simultaneous Multi-User Communication and Multi-Target Sensing}
\author{Kexin Chen, Yijie Mao,~\IEEEmembership{Member,~IEEE,} Longfei Yin,~\IEEEmembership{Student Member,~IEEE}, Chengcheng Xu, \\and  Yang Huang,~\IEEEmembership{Member,~IEEE}
\thanks{This work has been supported in part by the National Nature Science Foundation of China under Grant 62201347; and in part by Shanghai Sailing Program under Grant 22YF1428400. (\textit{Corresponding author: Yijie Mao})} 
\thanks{K. Chen and Y. Mao are with the School of Information Science and Technology, ShanghaiTech University, Shanghai 201210, China (email: chen.kx@shanghaitech.edu.cn; maoyj@shanghaitech.edu.cn).}
\thanks{L. Yin is with the Department of Electrical and Electronic Engineering, Imperial College London, London SW7 2AZ, U.K (email: longfei.yin17@imperial.ac.uk).}
\thanks{C. Xu is with the College of Electronic Engineering, National University of Defense Technology, Hefei 230037, China (email: xuchengcheng17@nudt.edu.cn).}
\thanks{Y. Huang is with the College of Electronic and Information Engineering, Nanjing University of Aeronautics and Astronautics, Nanjing 210016, China (email: yang.huang.ceie@nuaa.edu.cn).}	
\vspace{-1mm}
}

\maketitle

\begin{abstract}
In this paper, we initiate the study of rate-splitting multiple access (RSMA) for a mono-static integrated sensing and communication (ISAC) system, where the dual-functional base station (BS) simultaneously communicates with multiple users and detects multiple moving targets. 
We aim at optimizing the ISAC waveform to jointly maximize the max-min fairness (MMF)  rate of the communication users and minimize the largest eigenvalue of the Cramér-Rao bound (CRB) matrix for unbiased estimation. The CRB matrix considered in this work is general as it involves the estimation of the angular direction, the complex reflection coefficient, and the Doppler frequency for multiple moving targets.
Simulation results demonstrate that RSMA maintains a larger communication and sensing trade-off than conventional space-division multiple access (SDMA) and it is capable of detecting multiple targets with a high detection accuracy.
The finding highlights the potential of RSMA as an effective and powerful strategy for interference management in the general multi-user multi-target ISAC systems.
\end{abstract}


\section{Introduction}
Integrated sensing and communication (ISAC) has gained recognition as a promising enabling technology for 6G and beyond. 
The integration of these two functionalities into a unified framework is expected to unlock a wide range of new use cases and applications by elevating the overall performance, such as spectral efficiency and sensing accuracy.
One major challenge in ISAC lies in designing a dual-functional waveform, which is typically divided into the following three categories: sensing-centric design, communication-centric design, and joint design \cite{ISAC3/CRB}.
%
In this work, we dedicate to the joint design category, which has been a mainstream in ISAC.

\par 
Another challenge introduced by ISAC is the interference between communication and sensing, which occurs due to the overlapping frequency bands. 
Recent studies have shown that rate-splitting multiple access (RSMA), an advanced multiple access and dynamic interference management strategy, is capable to address the  challenge and achieves a flexible and robust interference management between communication and sensing or among communication users
\cite{RSMA-ISACcommonstream,RSMA-ISAC-partcsit,RSMA-ISACsatellite,RSMA-ISACpart2}.
The concept of RSMA-assisted ISAC was initially explored in\cite{RSMA-ISACcommonstream}, which highlights that the common stream of RSMA serves triple purposes of managing interference among communication users, managing interference between communication and radar, and beampattern approximation.
Further, \cite{RSMA-ISAC-partcsit} showed the advantages of RSMA-assisted ISAC in a practical scenario of partial channel state information (CSIT) and moving communication users. \cite{RSMA-ISACsatellite} extended RSMA-assisted ISAC from terrestrial communications to satellite communications. 
\cite{RSMA-ISACpart2} demonstrated that RSMA improves the trade-off between max-min fairness (MMF) rate and Cramér-Rao bound (CRB) of the single sensing target.
\par
In this work, we initiate the study of RSMA in a multi-user multi-target ISAC system. The main contributions of this work are listed as follows:
        \begin{itemize}
        \item We establish a novel and general RSMA-assisted mono-static ISAC system model that enables simultaneous communication with multiple users and detection of multiple targets. This is the first paper, to the best of our knowledge, that investigates RSMA in multi-user multi-target ISAC, as all previous works only focused on the single-target scenario.
        \item We derive a novel and general CRB radar sensing metric that is capable of measuring the angular direction, the complex reflection coefficient, and the Doppler frequency for multiple moving targets. This approach considers existing CRB metrics for single-target sensing or the detection of a limited set of parameters as special cases.
        \item We formulate and solve a non-convex optimization problem to optimize precoders and common rate allocations. The goal is to jointly maximize the MMF rate and minimize the largest eigenvalue of the derived general CRB metric. We first transform the original problem into an equivalent semidefinite programming (SDP) problem. Afterward, we use an iterative algorithm based on successive convex approximation (SCA) to solve this problem.
        \item Simulation results demonstrate, for the first time, that RSMA maintains a superior communication and sensing trade-off than SDMA for all targets, and it achieves high detection accuracy for multiple moving targets, as evaluated at the radar receiver using the Capon method.
        \end{itemize}
\par
\textit{Notations}: Matrices and vectors are represented by the boldface uppercase and lowercase letters, respectively. $\mathbf{1}^{M\times N}$ represents an all-one matrix with dimension $M\times N$. $\mathrm{Re}(\cdot)$ and $\mathrm{Im}(\cdot)$ respectively denote the real and imaginary parts of a complex scalar, vector, or matrix. $\mathrm{diag}(\mathbf{c})$ constructs a matrix with the entries of vector $\mathbf{c}$ placed along the diagonal. $\mathrm{tr}(\cdot)$, $(\cdot)^{T}$, $(\cdot)^{\ast}$, $(\cdot)^{H}$ and $(\cdot)^{-1}$ represent trace, transpose, conjugate, conjugate-transpose, and inverse opterations, respectively. $\left \| \cdot  \right \|_{2}$ denotes the Euclidean norm and $\odot $ signifies the Hadamard product.

\section{System Model} \label{model}
\begin{figure}[tb]
	\centerline{\includegraphics[width=0.3\textwidth]{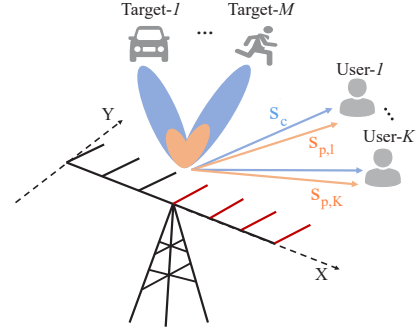}}
	\caption{The system model of the proposed mono-static RSMA-assisted ISAC.}
	\label{fig0}
\end{figure}

As shown in Fig. \ref{fig0}, we consider a mono-static ISAC system assisted by RSMA, where the transmit antennas are shared by the communication users and the moving targets.
The base station (BS), equipped with a uniform linear array (ULA) of 
$N_{t}$ transmit antennas and $N_{r}$ receive antennas, simultaneously communicates with $K$ single-antenna downlink communication users and detects $M$ moving targets. The communication users and moving targets are indexed by $\mathcal{K}=\left \{ 1,\dots ,K \right \}$ and $\mathcal{M}=\left \{ 1,\dots ,M \right \}$, respectively. 

\par
Consider the simplest and practical 1-layer RSMA model \cite{RSMAsurvey}, where the message $U_{k}$ intended for communication user $k$ is split into a common message $U_{c,k}$ and a private message $U_{p,k}$, respectively. All common parts are collectively encoded into a single common stream $s_{c}$, and the private parts are separately encoded into private streams $\left \{s_{p,k}\right \}_{k=1}^{K}$. 
Consider $N$ transmission and radar pulse blocks in one coherent processing interval (CPI) indexed by $\mathcal{N}=[1,\dots ,N]$, the transmit data stream vector at each time index $n$ is $\mathbf{s}[n]=[s_{c}[n],s_{p,1}[n],\dots,s_{p,K}[n]]^{T}\in \mathbb{C}^{(K+1)\times 1}$. 
Note that the common stream $s_{c}$ has demonstrated beneficial in ISAC systems by offering the triple functions of managing interference among communication users, managing interference between communication and radar, as well as beampattern approximation. RSMA with and without radar sequence show the same trade-off performance, thereby eliminating the need for an additional radar sequence in RSMA-assisted ISAC \cite{RSMA-ISACcommonstream}.
The streams are linearly precoded by the precoding matrix $ \mathbf{W}=[\mathbf{w}_{c},\mathbf{w}_{1},\dots, \mathbf{w}_{K}]\in \mathbb{C}^{N_{t}\times (K+1)}$, which remains consistent during one CPI. The transmit signal at time index $n$ is 
\begin{equation}
	\mathbf{x}[n]=\mathbf{W}\mathbf{s}[n]=\mathbf{w}_{c}s_{c}[n]+\sum_{k\in \mathcal{K}}\mathbf{w}_{k}s_{p,k}[n],
\end{equation}
where the data streams satisfy $ \mathbf{s}[n]\mathbf{s}[n]^{H}=\mathbf{I}_{K+1}$, implying that the entries are independent from each other. 
Hence, the covariance matrix of the transmit signal can be calculated by $\mathbf{R}_{x}=\frac{1}{N}\sum_{n\in \mathcal{N}}\mathbf{x}[n]\mathbf{x}[n]^{H}=\mathbf{W}\mathbf{W}^{H}$.

The signal received at the $k$th communication user at time index $n$ is given as
\begin{equation}
	\begin{aligned}
		y_{k}[n]&
		=\mathbf{h}_{k}^{H}\mathbf{x}[n]+z_{k}[n] \\
		&=\mathbf{h}_{k}^{H}\mathbf{w}_{c}s_{c}[n]+\sum_{i\in \mathcal{K}}\mathbf{h}_{k}^{H}\mathbf{w}_{i}s_{p,i}[n]+z_{k}[n], \forall k\in \mathcal{K},
	\end{aligned}
\end{equation}
where $\mathbf{h}_{k}\in \mathbb{C}^{N_{t}\times 1}$ is the communication channel between the BS and user $k$. It is assumed to be perfectly known at the BS and communication users. $z_{k}[n]$ is the additive white Gaussian noise (AWGN) received at  user $k$, which follows the distribution of $\mathcal{CN}(0,\sigma _{c}^{2})$. 

\par
As the transmit signal is also utilized for detecting the moving targets, the radar echo signal received at the BS at time index $n$ is defined as
\begin{equation} \label{Ys}
		\mathbf{y}_{s}[n]=\sum_{m\in \mathcal{M}}\alpha _{m}e^{j2\pi \mathcal{F}_{D_{m}}nT}\mathbf{b}(\theta _{m})\mathbf{a}^{T}(\theta _{m})\mathbf{x}[n]+\mathbf{z}_{s}[n],
\end{equation}
where $\left \{ \alpha _{m} \right \}_{m=1}^{M}$ represent the complex reflection coefficients proportional to the targets' radar cross-section (RCS). $\left \{ \mathcal{F}_{D_{m}} \right \}_{m=1}^{M}$ are the Doppler frequencies for different targets with $\mathcal{F}_{D_{m}}=\frac{2v_{m}f_{c}}{c}$, where $v_{m}$ denotes the velocity of the $m$th moving target and $c$, $f_{c}$ are the   speed of light and carrier frequency, respectively. $T$ represents the symbol period.
$\left \{ \theta_{m} \right \}_{m=1}^{M}$ denote the interested targets' direction of departure (DoD) as well as the direction of arrival (DoA), which are equal in a mono-static system. $\mathbf{a}(\theta_{m})=[1,e^{j\pi\sin (\theta_{m})},\dots,e^{j\pi(N_{t}-1)\sin (\theta_{m})} ]^{T}\in \mathbb{C}^{N_{t}\times 1}$, $\forall m \in \mathcal{M}$ is the transmit steering vector, where the distance between adjacent array elements is half-wavelength. And $\mathbf{b}(\theta_{m})\in \mathbb{C}^{N_{r}\times 1}$ denotes the receive steering vector defined in the same way as $\mathbf{a}(\theta_{m})$. 
$\mathbf{z}_{s}[n] \in \mathbb{C}^{N_{r}\times 1}$ denotes the AWGN following $\mathcal{CN}(\mathbf{0}^{N_{r}\times 1},\mathbf{Q})$, where $\mathbf{Q}=\sigma _{s}^{2}\mathbf{I}_{N_{r}}$.
\par
For notation simplicity, equation (\ref{Ys}) is equivalently rewritten as
\begin{equation}
	\mathbf{y}_{s}[n]=\mathbf{B}\mathbf{U}\mathbf{E}[n]\mathbf{A}^{T}\mathbf{x}[n]+\mathbf{z}_{s}[n],
\end{equation}
where
\begin{equation}\label{matrix for simplicity}
	\begin{aligned}
		&\mathbf{A}=[\mathbf{a}(\theta _{1}),\dots,\mathbf{a}(\theta _{M})], 
		\mathbf{B}=[\mathbf{b}(\theta _{1}),\dots,\mathbf{b}(\theta _{M})], \\
		&\boldsymbol{\alpha}=[\alpha _{1},\dots,\alpha _{M}]^{T}, 
		\boldsymbol{\theta }=[\theta _{1},\dots ,\theta _{M}]^{T}, 
		\mathbf{U}=\mathrm{diag}(\boldsymbol{\alpha}), \\
		&\mathbf{E}[n]=\mathrm{diag}([e^{j2\pi \mathcal{F}_{D_{1}}nT}, \dots, e^{j2\pi 			\mathcal{F}_{D_{M}}nT}]^{T}). 
	\end{aligned}
\end{equation}

\par
\section{Performance Metrics and Problem Formulation}\label{metric}
In this section, we specify the respective performance metrics for communication and radar sensing, namely, the MMF rate and the CRB. The corresponding optimization problem to jointly maximize the MMF rate and minimize the largest eigenvalue of CRB is then presented.
\subsection{Metrics for Multi-User Communication}
We select MMF rate to evaluate the communication performance in the considered multi-user multi-target ISAC system. Following the principle of 1-layer RSMA \cite{RSMAsurvey}, each user sequentially decodes the common stream and its own private stream. The corresponding achievable rates for the common and private streams at each user are given as
\begin{equation}
	R_{c,k}=\log_{2}\left(1+\frac{|\mathbf{h}_{k}^{H}\mathbf{w}_{c}|^{2}}{\sum_{i\in \mathcal{K}}|\mathbf{h}_{k}^{H}\mathbf{w}_{i}|^{2}+\sigma_{c}^{2}}\right), \forall k\in\mathcal{K},
\end{equation}
\begin{equation}
	R_{p,k}=\log_{2}\left(1+\frac{|\mathbf{h}_{k}^{H}\mathbf{w}_{k}|^{2}}{\sum_{i\in \mathcal{K},i\neq k}|\mathbf{h}_{k}^{H}\mathbf{w}_{i}|^{2}+\sigma_{c}^{2}}\right), \forall k\in\mathcal{K}.
\end{equation}
%
In order for each user to successfully decode the common stream, the achievable common rate is defined by $R_{c}=\min_{k\in \mathcal{K}}\{R_{c,k}\}=\sum_{k\in \mathcal{K}}C_{k}$, with $C_{k}$ denoting the allocated rate for transmitting user $k$'s common message. Therefore, the achievable rate of user $k$ is $R_{tot,k}=C_{k}+R_{p,k}$, $\forall k\in\mathcal{K}$, of which the minimum value $\mathop{\min}_{k\in \mathcal{K}}\{R_{tot,k}\}$ is  the MMF rate. 
\subsection{Metrics for Multi-Target Sensing}\label{sensing metric}
We choose the commonly used radar sensing metric CRB for target estimation. It is the lower bound on the variance of unbiased estimators\cite{ISAC3/CRB} and equivalent to the inverse of the fisher information matrix (FIM) denoted by $\mathbf{F}$, which means $\mathbf{CRB}=\mathbf{F}^{-1}$.
The matrix $\mathbf{F}$ involves the estimation of the angular direction, the complex reflection coefficient, and the Doppler frequency for multiple moving targets. With 
the parameters defined as $\mathbf{\xi }_{m}=\left \{ \theta_{m} ,\mathrm{Re}(\alpha _{m}),\mathrm{Im}(\alpha _{m}) ,\mathcal{F}_{D_{m}} \right \}^{T}$, $\forall m\in\mathcal{M}$, the FIM matrix $\mathbf{F}\in\mathbb{R}^{4M\times 4M}$ is given as
\begin{equation}\label{fisher}
	\mathbf{F}=2\begin{bmatrix}
		\mathrm{Re}(\mathbf{F}_{11}) & \mathrm{Re}(\mathbf{F}_{12}) & -\mathrm{Im}(\mathbf{F}_{12}) & -\mathrm{Im}(\mathbf{F}_{14}) \\ 
		\mathrm{Re}^{T}(\mathbf{F}_{12}) & \mathrm{Re}(\mathbf{F}_{22}) & -\mathrm{Im}(\mathbf{F}_{22}) & -\mathrm{Im}(\mathbf{F}_{24})\\ 
		-\mathrm{Im}^{T}(\mathbf{F}_{12}) & -\mathrm{Im}^{T}(\mathbf{F}_{22}) & \mathrm{Re}(\mathbf{F}_{22}) & \mathrm{Re}(\mathbf{F}_{24})\\ 
		-\mathrm{Im}^{T}(\mathbf{F}_{14}) & -\mathrm{Im}^{T}(\mathbf{F}_{24}) & \mathrm{Re}^{T}(\mathbf{F}_{24}) & \mathrm{Re}(\mathbf{F}_{44})
	\end{bmatrix}
\end{equation}
with $\mathbf{F}_{pq}$, $p,q\in\left \{ 1,2,4 \right \}$ being calculated as
\begin{equation}\label{eq:f}
	\begin{aligned}
		\mathbf{F}_{11}&=(\mathbf{\dot{B}}^{H}\mathbf{Q}^{-1}\mathbf{\dot{B}})\odot (\mathbf{U}^{\ast }\mathbf{A}^{H}\mathbf{R}_{x}^{\ast }\mathbf{A}\mathbf{U})\odot \mathbf{\Sigma }_{1}  \\
		&+(\mathbf{\dot{B}}^{H}\mathbf{Q}^{-1}\mathbf{B})\odot (\mathbf{U}^{\ast }\mathbf{A}^{H}\mathbf{R}_{x}^{\ast }\mathbf{\dot{A}}\mathbf{U})\odot \mathbf{\Sigma }_{1}  \\
		&+(\mathbf{B}^{H}\mathbf{Q}^{-1}\mathbf{\dot{B}})\odot (\mathbf{U}^{\ast }\mathbf{\dot{A}}^{H}\mathbf{R}_{x}^{\ast }\mathbf{A}\mathbf{U})\odot \mathbf{\Sigma }_{1}  \\
		&+(\mathbf{B}^{H}\mathbf{Q}^{-1}\mathbf{B})\odot (\mathbf{U}^{\ast }\mathbf{\dot{A}}^{H}\mathbf{R}_{x}^{\ast }\mathbf{\dot{A}}\mathbf{U})\odot \mathbf{\Sigma }_{1} , \\
		\mathbf{F}_{12}&=(\mathbf{\dot{B}}^{H}\mathbf{Q}^{-1}\mathbf{B})\odot (\mathbf{U}^{\ast }\mathbf{A}^{H}\mathbf{R}_{x}^{\ast }\mathbf{A})\odot \mathbf{\Sigma }_{1}  \\
		&+(\mathbf{B}^{H}\mathbf{Q}^{-1}\mathbf{B})\odot (\mathbf{U}^{\ast }\mathbf{\dot{A}}^{H}\mathbf{R}_{x}^{\ast }\mathbf{A})\odot \mathbf{\Sigma }_{1}, \\
		\mathbf{F}_{14}&=(\mathbf{\dot{B}}^{H}\mathbf{Q}^{-1}\mathbf{B})\odot (\mathbf{U}^{\ast }\mathbf{A}^{H}\mathbf{R}_{x}^{\ast }\mathbf{A}\mathbf{U})\odot \mathbf{\Sigma }_{2}  \\
		&+(\mathbf{B}^{H}\mathbf{Q}^{-1}\mathbf{B})\odot (\mathbf{U}^{\ast }\mathbf{\dot{A}}^{H}\mathbf{R}_{x}^{\ast }\mathbf{A}\mathbf{U})\odot \mathbf{\Sigma }_{2}, \\
		\mathbf{F}_{22}&=(\mathbf{B}^{H}\mathbf{Q}^{-1}\mathbf{B})\odot (\mathbf{A}^{H}\mathbf{R}_{x}^{\ast }\mathbf{A})\odot \mathbf{\Sigma }_{1}, \\
		\mathbf{F}_{24}&=(\mathbf{B}^{H}\mathbf{Q}^{-1}\mathbf{B})\odot (\mathbf{A}^{H}\mathbf{R}_{x}^{\ast }\mathbf{A}\mathbf{U})\odot \mathbf{\Sigma }_{2}, \\
		\mathbf{F}_{44}&=(\mathbf{B}^{H}\mathbf{Q}^{-1}\mathbf{B})\odot (\mathbf{U}^{\ast }\mathbf{A}^{H}\mathbf{R}_{x}^{\ast }\mathbf{A}\mathbf{U})\odot \mathbf{\Sigma }_{3}, 
	\end{aligned}
\end{equation}		
where 
\begin{equation}
	\label{eq:fii}
	\resizebox{0.89\hsize}{!}{$
	\begin{aligned}
	&	\mathbf{\dot{A}}=\begin{bmatrix}
			\frac{\partial \mathbf{a}(\theta _{1})}{\partial \theta _{1}}&,\dots, & \frac{\partial \mathbf{a}(\theta _{M})}{\partial \theta _{M}} 
		\end{bmatrix} , \,
		\mathbf{\dot{B}}=\begin{bmatrix}
			\frac{\partial \mathbf{b}(\theta _{1})}{\partial \theta _{1}},\dots,  \frac{\partial \mathbf{b}(\theta _{M})}{\partial \theta _{M}} 
		\end{bmatrix},\\
	&	(\mathbf{\Sigma}_{1})_{ij}=\sum_{n\in\mathcal{N}}e^{j2\pi (\mathcal{F}_{D_{j}}-\mathcal{F}_{D_{i}})nT}, \forall i,j\in \mathcal{M},  \\
	&	(\mathbf{\Sigma}_{2})_{ij}=\sum_{n\in\mathcal{N}}2\pi nTe^{j2\pi (\mathcal{F}_{D_{j}}-\mathcal{F}_{D_{i}})nT}, \forall i,j\in  \mathcal{M}, \\
	&	(\mathbf{\Sigma}_{3})_{ij}=\sum_{n\in\mathcal{N}}(2\pi nT)^{2}e^{j2\pi (\mathcal{F}_{D_{j}}-\mathcal{F}_{D_{i}})nT}, \forall i,j\in \mathcal{M},
	\end{aligned}
		$}
\end{equation}
and $\mathbf{A}$, $\mathbf{B}$, $\mathbf{U}$ are specified in (\ref{matrix for simplicity}). $\mathbf{R}_{x}=\mathbf{W}\mathbf{W}^{H}$ and $\mathbf{Q}=\sigma _{s}^{2}\mathbf{I}_{N_{r}}$ are the respective covariance matrices of the transmit signal and the AWGN at the BS. Note that $\mathbf{Q}^{-1}=\frac{1}{\sigma _{s}^{2}}\mathbf{I}_{N_{r}}$. The detailed derivation procedure is provided in the Appendix.
\par
\textit{Remark}: The CRB metric derived in this work is more general than the one considered in existing works \cite{RSMA-ISACpart2, mutitarCRB2008}.
On the one hand, the FIM in (\ref{fisher}) involves the estimation of multiple moving targets, which considers the single-target FIM in \cite{RSMA-ISACpart2} as a special case. On the other hand, the FIM in (\ref{fisher})  measures the Doppler frequency in the radar echo signal. This contrasts with \cite{mutitarCRB2008}, which considers multiple targets but neglects the Doppler frequency, i.e, $(\mathbf{\Sigma}_{1})_{ij}=L$. Moreover, in contrast to observations made for the single-target scenario in \cite{RSMA-ISACpart2}, we note that the velocities of multiple moving targets indeed impact the trade-off performance between the CRB and MMF rate. According to equations (\ref{fisher})--(\ref{eq:fii}), it becomes apparent that the CRB associated with the velocities of multiple moving targets depends on the disparities in their Doppler frequencies, denoted as $(\mathcal{F}_{D_{j}}-\mathcal{F}_{D_{i}}), \forall i,j\in \mathcal{M}$.

\subsection{Problem Formulation}
In this work, we aim at jointly maximizing the MMF rate of the communication users and minimizing the largest eigenvalue of the CRB matrix of multiple moving targets. Note that the latter is equivalent to maximizing the smallest eigenvalue of $\mathbf{F}$\cite{mutitarCRB2008}. The ISAC waveform optimization problem is

\begin{subequations}\label{p1}
	\begin{align}
&\mathop{\max}_{\mathbf{c},\mathbf{W},g}\,\,\, \mathop{\min}_{k\in \mathcal{K}}\{R_{tot,k}\}+\mu g \\
		s.t.\,\,\,\,&\mathbf{F}\succeq  g\mathbf{I}_{4M}, \label{opt 1}\\
		&\mathbf{c}\geq \mathbf{0}, \label{opt 2} \\
		&R_{c,k}\geq \sum_{i\in \mathcal{K}}C_{i}, \forall k\in \mathcal{K}, \label{opt 3}\\
		&\mathrm{diag}(\mathbf{W}\mathbf{W}^{H})=\frac{P\mathbf{1}^{N_{t}\times 1}}{N_{t}}, \label{opt 4}
	\end{align}
\end{subequations}
where $\mathbf{c}=\left[C_{1},\dots,C_{K}\right]^{T}\in\mathbb{C}^{K\times 1} $ is the common rate allocation among users.  $\mu$ is the regularization parameter, through which we can shift the priority between communication and sensing functionality. $\mathbf{I}_{4M}$ is the identity matrix a dimension of $4M\times 4M$. Constraint (\ref{opt 1}) guarantees the matrix ($\mathbf{F}-g\mathbf{I}_{4M}$) is positive semi-definite, and $g$ is the auxiliary value corresponding to the smallest eigenvalue of $\mathbf{F}$. (\ref{opt 2}) guarantees the non-negativity of the common rate allocations. Constraint (\ref{opt 3}) ensures that each communication user can decode the common stream successfully. The power at the BS is constraint by (\ref{opt 4}), with $P$ denoting the total power budget\cite{RSMA-ISACpart2}. 

Compared to the problem for the single-target case in \cite{RSMA-ISACpart2}, the interference between communication and sensing becomes more intricate when dealing with multiple targets in (\ref{p1}). Moreover, $\mathbf{F}$ in (\ref{opt 1}) confronts with a substantially larger matrix dimension compared to  the single-target case, expanding from $4\times4$ to $4M\times 4M$. It also leads to a natural complication in factors influencing estimation performance, such as differences in angular direction and Doppler frequencies.
\section{ Optimization Framework}\label{formu}
In this section, we present the optimization framework designed to solve problem (\ref{p1}). 
\par
By defining $\mathbf{H}_{k}=\mathbf{h}_{k}\mathbf{h}_{k}^{H}$, $\mathbf{W}_{c}=\mathbf{w}_{c}\mathbf{w}_{c}^{H}$, and $\mathbf{W}_{k}=\mathbf{w}_{k}\mathbf{w}_{k}^{H}, \forall k \in \mathcal{K}$ , where $\mathrm{rank}\left(\mathbf{W}_{c}\right)=1$ and $\mathrm{rank}(\mathbf{W}_{k})=1$, we have  $\mathbf{R}_{x}=\mathbf{W}\mathbf{W}^{H}=\mathbf{W}_{c}+\sum_{k\in \mathcal{K}}\mathbf{W}_{k}$. Problem (\ref{p1}) is then equivalently transformed into a SDP problem.
To deal with the non-convex rate expressions, we introduce auxiliary variables $r_{m}$ and $\mathbf{r}=\left[r_{p,1},\dots,r_{p,K}\right]^{T}\in\mathbb{C}^{K\times1}$, where the former is the MMF rate of communication users and the latter denotes the lower bounds of corresponding private rate $\{R_{p,k}\}_{k=1}^{K}$.
We also introduce slack variables $\{\varphi _{i,k}\}_{k=1}^{K}$, $\{\delta _{i,k}\}_{k=1}^{K}$, $i\in\{c, p\}$, where $\{\mathrm{e}^{\varphi _{i,k}}\}_{k=1}^{K}$, $i\in\{c, p\}$ are the upper bounds of the interference-plus-noise term in common and private rate expressions, respectively. Correspondingly, $\{\mathrm{e}^{\delta _{i,k}}\}_{k=1}^{K}$, $i\in\{c, p\}$ are the lower bounds of the signal term plus interference-plus-noise term in common and private rate expressions. 
Considering the high computational complexity caused by the nonlinear expressions $\{\mathrm{e}^{\delta _{i,k}}\}_{k=1}^{K}$, $i\in\{c, p\}$, we utilize slack variables $\{\zeta_{i,k}\}_{k=1}^{K}$, $i\in\{c, p\}$ to denote the upper bounds of them. With all the aforementioned slack variables, problem (\ref{p1}) can be equivalently transformed as

\begin{subequations}\label{p2}
	\begin{align}
		(\mathcal{P}_2)\,\,&\mathop{\max}_{\mathbf{c}, \mathbf{W}_{c}, \{\mathbf{W}_{k}\}_{k=1}^{K}, \mathbf{r}, r_{m}, g, \delta, \varphi, \zeta}\,\,\,r_{m}+\mu g\\
		s.t.\,\,\,\,&\mathrm{diag}(\mathbf{W}_{c}+\sum_{k\in \mathcal{K}}\mathbf{W}_{k})=\frac{P\mathbf{1}^{N_{t}\times 1}}{N_{t}}, \forall k \in \mathcal{K},\label{p2_power}\\
		&\mathbf{W}_{c}\succeq \mathbf{0}, \mathbf{W}_{k}\succeq \mathbf{0}, \forall k \in \mathcal{K},\label{p2_p0}\\
		&\mathrm{rank}(\mathbf{W}_{c})=1, \mathrm{rank}(\mathbf{W}_{k})=1, \forall k \in \mathcal{K}, \label{opt_rank1}\\		
		&C_{k}+r_{p,k}\geq r_{m}, \forall k \in \mathcal{K},\label{p2_rth}\\
		&\sum_{j\in \mathcal{K}}C_{j} \mathrm{ln2}\leq \delta _{c,k}-\varphi _{c,k}, \forall k \in \mathcal{K},\label{opt_common}\\
		&r_{p,k} \mathrm{ln2}\leq \delta _{p,k}-\varphi _{p,k}, \forall k \in \mathcal{K},\label{opt_pri}\\
		&\mathrm{e}^{\varphi _{c,k}}\geq \sum_{i\in \mathcal{K}}\mathrm{tr}(\mathbf{W}_{i}\mathbf{H}_{k})+\sigma_{c}^{2}, \forall k \in \mathcal{K},\label{opt_common1}\\
		&\mathrm{e}^{\varphi _{p,k}}\geq \sum_{i\in \mathcal{K},i\neq k}\mathrm{tr}(\mathbf{W}_{i}\mathbf{H}_{k})+\sigma_{c}^{2}, \forall k \in \mathcal{K},\label{opt_pri1}\\
		&\zeta_{i,k}\mathrm{ln}\zeta_{i,k}\geq \delta _{i,k}\zeta_{i,k},  i\in\{c, p\}, \forall k \in \mathcal{K},\label{opt_common_pri3}\\
		\zeta_{c,k}&\leq \sum_{i\in \mathcal{K}}\mathrm{tr}(\mathbf{W}_{i}\mathbf{H}_{k})+\mathrm{tr}(\mathbf{W}_{c}\mathbf{H}_{k})+\sigma _{c}^{2},  \forall k \in \mathcal{K},\label{opt_common2}\\
		\zeta_{p,k}&\leq \sum_{i\in \mathcal{K}}\mathrm{tr}(\mathbf{W}_{i}\mathbf{H}_{k})+\sigma_{c}^{2}, \forall k \in \mathcal{K},\label{opt_pri2}\\
		&\textrm{(\ref{opt 1}), (\ref{opt 2})}.
	\end{align}
\end{subequations}
\par
The transformed problem (\ref{p2}) remains non-convex due to constraints (\ref{opt_common1})-(\ref{opt_common_pri3}). We then approximate the convex left hand sides $\mathrm{e}^{\varphi _{i,k}}$, $\zeta_{i,k}\mathrm{ln}\zeta_{i,k}$, $i\in\{c, p\}$, $\forall k \in \mathcal{K}$ by respectively employing the first-order Taylor approximation at points $\varphi _{i,k}^{[t]}$, $\zeta _{i,k}^{[t]}$, $i\in\{c, p\}$, $\forall k \in \mathcal{K}$ at each iteration $t$.
 Constraints (\ref{opt_common1})-(\ref{opt_common_pri3}) are therefore approximated at iteration $t$ as
%
\begin{equation}\label{opt_tay1}
	(1+\varphi _{c,k}-\varphi _{c,k}^{[t]})\mathrm{e}^{\varphi _{c,k}^{[t]}}\geq  \sum_{i\in \mathcal{K}}\mathrm{tr}(\mathbf{W}_{i}\mathbf{H}_{k})+\sigma_{c} ^{2}, \forall k \in \mathcal{K},
\end{equation}
\begin{equation}\label{opt_tay2}
	(1+\varphi _{p,k}-\varphi _{p,k}^{[t]})\mathrm{e}^{\varphi _{p,k}^{[t]}}\geq  \sum_{i\in \mathcal{K},i\neq k}\mathrm{tr}(\mathbf{W}_{i}\mathbf{H}_{k})+\sigma_{c}^{2},
\forall k \in \mathcal{K},
\end{equation}
\begin{equation}\label{opt_soc}
	\begin{aligned}
		&\left \| \Big [2\sqrt{\zeta_{i,k}^{[t]}}, \delta _{i,k}+\zeta _{i,k}-(1+\mathrm{ln}\zeta_{i,k}^{[t]})\Big ]\right \|_{2}\\
		&\leq-\delta _{i,k}+\zeta _{i,k}+(1+\mathrm{ln}\zeta_{i,k}^{[t]}), i\in\{c, p\}, \forall k \in \mathcal{K},
	\end{aligned}
\end{equation}
\par
To handle the rank-one constraints (\ref{opt_rank1}), we define $\mathbf{u}_{c,max}^{[t]}$ and $\{\mathbf{u}_{k,max}^{[t]}\}_{k=1}^{K}$ as the normalized eigenvectors related to the maximum eigenvalues of $\mathbf{W}_{c}^{[t]}$ and $\{\mathbf{W}_{k}^{[t]}\}_{k=1}^{K}$, through which we then move (\ref{opt_rank1}) to the objective function by
\begin{equation}\label{crank}
	\begin{aligned}
C^{[t]}_{\mathrm{rank}}=&\rho \big\{ \big[\mathrm{tr}(\mathbf{W}_{c})-(\mathbf{u}_{c,max}^{[t]})^{H}\mathbf{W}_{c}\mathbf{u}_{c,max}^{[t]}\big]\\		+&\sum_{k\in \mathcal{K}}\big[\mathrm{tr}(\mathbf{W}_{k})-(\mathbf{u}_{k,max}^{[t]})^{H}\mathbf{W}_{k}\mathbf{u}_{k,max}^{[t]}\big] \big \},
	\end{aligned}
\end{equation}
where $\rho$ is a negative penalty constant.
\par
Based on the aforementioned approximation, we solve problem (11) via a sequence of convex subproblems. At iteration $t$, we solve the following subproblem based on the optimal solution $\mathbf{W}_{c}^{[t-1]}$, $\{\mathbf{W}_{k}^{[t-1]}\}_{k=1}^{K}$, $\delta^{[t-1]}$, $\varphi^{[t-1]}$, $\zeta^{[t-1]}$ obtained from the previous iteration:

\begin{subequations}\label{p3}
	\begin{align}
		(\mathcal{P}_3)\,\,&\mathop{\max}_{\mathbf{c}, \mathbf{W}_{c}, \{\mathbf{W}_{k}\}_{k=1}^{K}, \mathbf{r}, r_{m}, g, \delta, \varphi, \zeta}\,\,\,r_{m}+\mu g+C^{[t]}_{\mathrm{rank}} \\
		s.t.\,\,\,\,&\textrm{(\ref{opt 1}), (\ref{opt 2}), (\ref{p2_power}), (\ref{p2_p0}), (\ref{p2_rth})-(\ref{opt_pri}), (\ref{opt_common2}), }\nonumber\\
		&\textrm{(\ref{opt_pri2}), (\ref{opt_tay1})-(\ref{opt_soc}),}
	\end{align}
\end{subequations}
\begin{algorithm}[t]
	\textbf{Initialize}: $t\leftarrow0$, $\mathbf{W}_{c}^{[t]}$, $\{\mathbf{W}_{k}^{[t]}\}_{k=1}^{K}$, $\delta^{[t]}$, $\varphi^{[t]}$, $\zeta^{[t]}$, $\mathrm{Opt}^{[t]}$ \;
	\Repeat{$|\mathrm{Opt}^{[t]}-\mathrm{Opt}^{[t-1]}|<\tau$}{
		$t\leftarrow t+1$\;
		Solve problem (\ref{p3}) based on $\mathbf{W}_{c}^{[t-1]}$, $\{\mathbf{W}_{k}^{[t-1]}\}_{k=1}^{K}$, $\delta^{[t-1]}$, $\varphi^{[t-1]}$, $\zeta^{[t-1]}$ and denote the optimal value by $\mathbf{W}_{c}^{\star}$, $\{\mathbf{W}_{k}^{\star}\}_{k=1}^{K}$, $\delta^{\star}$, $\varphi^{\star}$, $\zeta^{\star}$ and the objective function by $\mathrm{Opt}^{\star}$\;
		Update $\mathbf{W}_{c}^{[t]}\leftarrow \mathbf{W}_{c}^{\star}$,  $\{\mathbf{W}_{k}^{[t]}\}_{k=1}^{K}\leftarrow\{\mathbf{W}_{k}^{\star}\}_{k=1}^{K}$, $\delta^{[t]}\leftarrow\delta^{\star}$, $\varphi^{[t]}\leftarrow\varphi^{\star}$, $\zeta^{[t]}\leftarrow\zeta^{\star}$, $\mathrm{Opt}^{[t]}\leftarrow\mathrm{Opt}^{\star}$\;			 	}	
	\caption{ISAC waveform optimization algorithm }
	\label{algor}
\end{algorithm}
\vspace{-4mm}
\par
The optimization procedure to solve problem (\ref{p1}) is detailed in Algorithm 1, with $\tau$ denoting the error tolerance value. The solution of problem (\ref{p3}) at iteration $t-1$ is a feasible solution at iteration $t$, and the objective function is monotonically increasing. As the objective function is bounded by the transmit power constraint, we then guarantee the convergence of problem (\ref{p3}). 
Interior-point methods are employed to address problem (\ref{p3}), which only involves linear matrix inequality (LMI) and SOC constraints. We then obtain the worst-case computational complexity of Algorithm 1 as $\mathcal{O} ([N_{t}^{2}(K+1)]^{3.5}\mathrm{log}(1/\tau) )$.
\section{Numerical Results}\label{result}
In this section, we demonstrate the performance of the proposed RSMA-assisted multi-user multi-target ISAC model. The respective performance metrics for communication  and sensing are the MMF rate and the trace of the weighted CRB, which is expressed as $\mathrm{tr}(\mathbf{\Lambda}\mathbf{F}^{-1})$, where $\mathbf{\Lambda}=\mathrm{diag}([\lambda _{1},\dots,\lambda _{4M}]^{T})\in\mathbb{C}^{4M\times4M}$ denotes the weights of different target parameters. Unless otherwise specified, the weights are set as $\left \{ \lambda _{i} \right\} _{i=1}^{4M}=1$\cite{bi4para}. When varying numbers of targets are considered, for fairness of comparison, we consider the sensing performance metric as the average trace of CRB defined by $\mathrm{tr}(\mathbf{F}^{-1})/M$.

Unless otherwise specified, we consider the scenario where the BS is equipped with  $N_{t}=4$ transmit antennas and $N_{r}=9$ receive antennas and it serves $K=4$ communication users. The total power budget at the BS is $P=20\ \mathrm{dBm}$ and we assume $\sigma_{c}^{2}=0\ \mathrm{dBm}$ without loss of generality. The SNR of radar is calculated by $\mathrm{SNR}_{radar}=|\alpha |P/\sigma _{s}^{2}=-20\ \mathrm{dBm}$, where $|\alpha |=|\{\alpha_{m}\}_{m=1}^{M}|=1, \forall m\in \mathcal{M}$. Consider $N=1024$ radar pulse blocks in one CPI.
The channels between the BS and the communication users are assumed to be Rayleigh fading with each entry following the complex Gaussian distribution as $\mathcal{CN}(0,1)$. We consider 7 different targets, among which targets 1-3 are located at 45$^{\circ}$, 30$^{\circ}$, 15$^{\circ}$ with velocities of 10 $\mathrm{m/s}$, 14 $\mathrm{m/s}$, 18 $\mathrm{m/s}$, respectively. Targets 4-7 are at 0$^{\circ}$, 34$^{\circ}$, 18$^{\circ}$, 9$^{\circ}$ with the same velocity of 10 $\mathrm{m/s}$. We use linearly precoded SDMA as a baseline, which is achieved through disabling the common stream of RSMA. The results are averaged over 100 channel realizations. 
\par

\begin{figure}[t]
	\centering
    \includegraphics[width=0.4\textwidth]{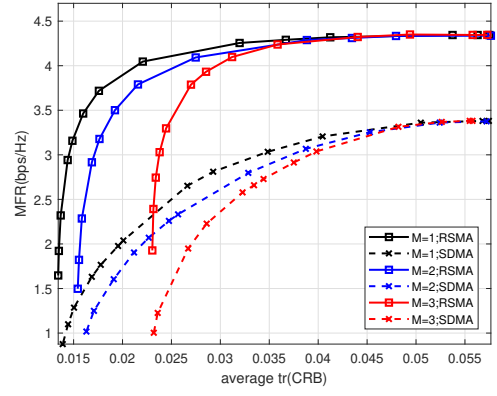}
	\caption{The trade-off performance under different numbers of targets $M=1-3$. $N_{t}=4$, $N_{r}=9$, $K=4$, $N=1024$, $\mathrm{SNR}_{radar}=-20\ \mathrm{dBm}$.}
	\label{fig1}
 \vspace{-4mm}
\end{figure}

Fig. \ref{fig1} illustrates the trade-off performance between communication and sensing when $M=1$ (target 1), $M=2$ (target 1-2) and $M=3$ (target 1-3). We observe that RSMA exhibits an explicit trade-off region gain over SDMA. Attributing to the additional degree-of-freedom (DoF) introduced by the common stream, RSMA achieves a superior MMF rate than SDMA at the rightmost corner point. As the number of targets increases, the trade-off regions of both RSMA and SDMA become worse since the loss of the beamforming power at each single target. Surprisingly, we observe that RSMA is capable of detecting more targets than SDMA while maintaining the QoS of the communication users, thus showing the great potential of RSMA to enhance the sensing functionality.
\par
\begin{figure}[t]
	\centering
    \includegraphics[width=0.4\textwidth]{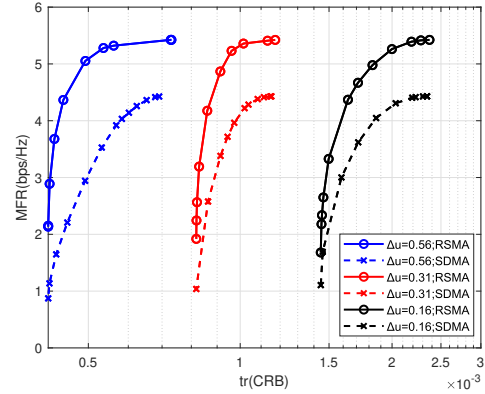}
	\caption{The trade-off performance under different angle difference. $N_{t}=4$, $N_{r}=5$, $K=4$, $M=2$, $N=256$, $\mathrm{SNR}_{radar}=10\ \mathrm{dBm}$.}
	\label{fig2}
 \vspace{-4mm}
\end{figure}

Fig. \ref{fig2} shows the trade-off comparison between RSMA and SDMA when the angle difference between $M=2$ targets defined by $\Delta u=\sin(\theta_{1})-\sin(\theta_{2}) $ varies. To explicitly investigate the influence of the angle difference between the target, $\Delta f=\mathcal{F}_{D_{1}}-\mathcal{F}_{D_{2}}=0$ is considered. There are $N_{t}=4$ transmit antennas and $N_{r}=5$ receive antennas at the BS. We set $\mathrm{SNR}_{radar}=10\  \mathrm{dBm}$, $N=256$ and generate $K=4$  communication user channels randomly by following the complex Gaussian distribution\cite{RSMA-ISACcommonstream}. As the angle difference grows from $ \Delta u=0.16$ (target 7, 4), $ \Delta u=0.31$ (target 6, 4) to $ \Delta u=0.56$ (target 5, 4), the sensing metric $\mathrm{tr(CRB)}$  tends to become lower, owing to the decrease of the interference between the radar echo signals at the receiver. We observe that under various angle differences, the proposed RSMA consistently outperforms SDMA.

\begin{figure}[t]
	\centering
    \includegraphics[width=0.4\textwidth]{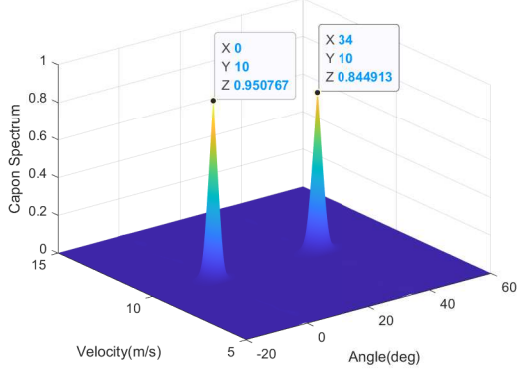}
	\caption{Capon estimation based on the RSMA-assisted ISAC beamforming.}
	\label{fig3}
 \vspace{-4mm}
\end{figure}

\begin{figure}[t]
	\centering
    \includegraphics[width=0.4\textwidth]{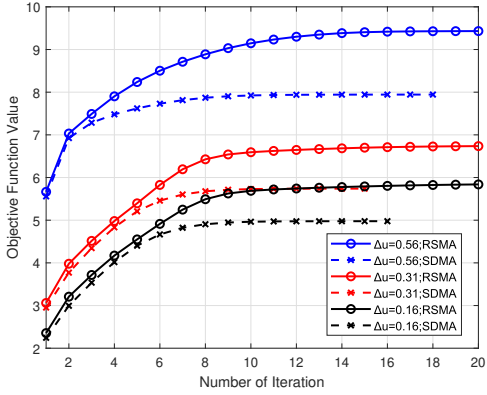}
	\caption{The convergence performance of Algorithm 1, $\mu=1e^{-3}$.}
	\label{objective}
\vspace{-4mm}
\end{figure}
\par
To further evaluate the sensing capability of the proposed ISAC system, we employ the Capon beamformer to estimate the angles and velocities of the targets, which maximizes SINR and reduces noise and interference while ensuring that the desired signal is not distorted. 
 The Capon beamformer $\mathbf{w}_{p} \in \mathbb{C}^{N_{r}\times 1}$ is expressed as
\begin{equation}\label{capon weight}
\mathbf{w}_{p}(\theta)=\frac{\mathbf{R}_{y}^{-1}\mathbf{b}(\theta )}{\mathbf{b}^{H}(\theta )\mathbf{R}_{y}^{-1}\mathbf{b}(\theta )},
\end{equation}
where $	\mathbf{R}_{y}=\frac{1}{N}\sum_{n\in\mathcal{N}}\mathbf{y}_{s}[n]\mathbf{y}_{s}[n]^{H}$ is the covariance of the received signal. The Capon estimation of the complex reflection coefficient is defined as the minimizer of the following cost-function
\begin{equation}\label{capon estimation}
	\begin{aligned}
		{\hat{\alpha }}(\theta, v)&=\mathop{\mathrm{argmin}}_{\alpha}E\big\{ |\mathbf{w}_{p}^{H}(\theta )\mathbf{y}_{s}[n]\\
		&-\mathbf{w}_{p}^{H}(\theta ) \alpha e^{j 2\pi \mathcal{F}_{D}nT}\mathbf{b}(\theta )\mathbf{a}^{T}(\theta)\mathbf{x}[n]|^{2} \big\}\\
		&=E\left \{ \frac{\mathbf{w}_{p}^{H}(\theta )\mathbf{y}_{s}[n]\mathbf{x}^{H}[n]\mathbf{a}^{\ast}(\theta ) e^{-j 2\pi \mathcal{F}_{D}nT}}{\mathbf{a}^{T}(\theta)\mathbf{x}[n]\mathbf{x}^{H}[n]\mathbf{a}^{\ast}(\theta )} \right \}.
	\end{aligned}
\end{equation}
Substituting (\ref{capon weight}) into (\ref{capon estimation}), we have
\begin{equation}\label{capon w}
	\begin{aligned}
		\hat{\alpha}(\theta, v)
	&=\frac{\mathbf{b}^{H}(\theta)\mathbf{R}_{y}^{-1}\mathbf{Y}_{s}\mathbf{D}\mathbf{X}^{H}\mathbf{a}^{\ast}(\theta)}{\mathbf{b}^{H}(\theta)\mathbf{R}_{y}^{-1}\mathbf{b}(\theta)p(\theta)N},
	\end{aligned}		
\end{equation}
where $p(\theta)=\mathbf{a}^{T}(\theta)\mathbf{R}_{x}\mathbf{a}^{\ast}(\theta)$ is the transmit beam pattern, 		$\mathbf{Y}_{s}=[\mathbf{y}_{s}[1],\dots,\mathbf{y}_{s}[N]]$, $\mathbf{X}=[\mathbf{x}[1],\dots,\mathbf{x}[N]]$, $\mathbf{D}=\mathrm{diag}([e^{-j 2\pi \mathcal{F}_{D}T},\dots,e^{-j 2\pi \mathcal{F}_{D}NT}]^{T})$. Obtaining $\hat{\alpha}$ at each grid point of $(\theta, v)$ requires two dimension calculation as \cite{caponestimation}
\begin{equation}
		\{\theta_{m}, v_{m}\}_{m=1}^{M}=\mathop{\mathrm{argmax}}_{\theta, v}|\hat{\alpha}(\theta, v)|^{2}.
\end{equation}
Therefore, we obtain $M$ peak points corresponding to the $M$ targets. Fig. \ref{fig3} shows two peak points, which correspond to targets 4 and 5 as specified in the simulation setting of Fig. \ref{fig2}. The precoding matrix used here is the optimized solution of (\ref{p3}) when $N_{t}=4$, $N_{r}=5$, $\mathrm{SNR}_{radar}=10\ \mathrm{dBm}$, $N=256$, and $\mu=1e^{-3}$. It is evident from Fig. \ref{fig3} that the parameters of the two moving targets are correctly estimated, implying the high detection accuracy of RSMA.

Additionally, we show the convergence performance of Algorithm 1 in Fig. \ref{objective} with $K = 4$ and $M = 2$. A negative penalty constant $\rho$ of $-0.1$  is employed to strike a balance between the convergence performance and the number of iterations required. It is clear that the convergence of the objective values within  the proposed SCA-based algorithm is guaranteed, with convergence typically achieved within twenty iterations for both RSMA and SDMA. 

\section{Conclusion} \label{conclude}
This work initiates the study of RSMA in a mono-static ISAC system with multiple communication users and multiple sensing targets.  We derive a  general CRB sensing metric which embraces the estimation of the angular direction, the complex reflection coefficient, and the Doppler frequency for multiple targets. By designing the transmit waveform to maximize the MMF rate of multiple users and minimize the largest eigenvalue of CRB of multiple moving targets, we show that RSMA achieves a better communication and sensing trade-off than conventional linearly precoded SDMA in multi-user multi-target ISAC. 
Additionally, the trade-off gain of RSMA grows with increasing angle difference between targets. 
Therefore, we conclude that RSMA offers a highly effective interference management solution. It has great potential for synergizing with ISAC in future wireless networks.
\section*{Appendix} 
\section*{ derivation of the FIM in (\ref{fisher}) } 
\label{Appendix}
CRB is the lower bound on the variance of unbiased estimators and is given by $\mathbf{CRB}=\mathbf{F}^{-1}$. The matrix $\mathbf{F}$ is related to four target parameters $\mathbf{\xi }_{m}=\left \{ \theta_{m} ,\mathrm{Re}(\alpha _{m}),\mathrm{Im}(\alpha _{m}) ,\mathcal{F}_{D_{m}} \right \}^{T}$. With definition of $\mathbf{v}[n]=\mathbf{y}_{s}[n]-\mathbf{z}_{s}[n]$, 
we note that
\begin{equation}\label{Fff}
\mathbf{F}_{\mathcal{F}_{D_{i}} \mathcal{F}_{D_{j}}}=2\mathrm{Re}\Big[\mathrm{tr}\Big\{ \sum_{n\in\mathcal{N}} \frac{\partial\mathbf{v}[n]^{H} }{\partial\mathcal{F}_{D_{i}} }\mathbf{Q}^{-1}\frac{\partial\mathbf{v}[n]}{\partial\mathcal{F}_{D_{j}}}\Big \}\Big], \forall i,j\in\mathcal{M},
\end{equation}
the partial derivative can be calculated as	\begin{equation}\label{derivate}
		\frac{\mathbf{v}[n]}{\partial \mathcal{F}_{D_{i}} }=\mathbf{B}\mathbf{U}\left ( j2\pi nT \right )\mathbf{E}[n]\mathbf{e}_{i}\mathbf{e}_{i}^{T}\mathbf{A}^{T}\mathbf{x}[n], \forall i\in\mathcal{M},
\end{equation}
where $\mathbf{e}_{i}$ is the $i$th column of $\mathbf{I}_{K}$. Since $\mathrm{tr}(\mathbf{A}\mathbf{B})=\mathrm{tr}(\mathbf{B}\mathbf{A})$ and $\mathbf{U}$, $\mathbf{E}[n]$ are diagonal matrices, (\ref{Fff}) can be rewritten as
\begin{equation}
	\begin{aligned}
		\mathbf{F}_{\mathcal{F}_{D_{i}} \mathcal{F}_{D_{j}}}&=2\mathrm{Re}\Big [\mathrm{tr}\Big \{ \sum_{n\in\mathcal{N}}\big ( \mathbf{B}\mathbf{U}\left ( j2\pi nT \right )\mathbf{E}[n]\mathbf{e}_{i}\mathbf{e}_{i}^{T}\mathbf{A}^{T}\times  \\
		& \mathbf{x}[n]\big )^{H}\mathbf{Q}^{-1}\left ( \mathbf{B}\mathbf{U}\left ( j2\pi nT \right )\mathbf{E}[n]\mathbf{e}_{j}\mathbf{e}_{j}^{T}\mathbf{A}^{T}\mathbf{x}[n]\right ) \Big \}\Big] \\
		&=2\mathrm{Re}\Big [\mathrm{tr}\Big \{
\sum_{n\in\mathcal{N}}\mathbf{e}_{i}^{T}\mathbf{B}^{H}\mathbf{Q}^{-1}\mathbf{B}\mathbf{e}_{j} \mathbf{e}_{j}^{T}\mathbf{U}\mathbf{E}[n]j2\pi nT \\
		&\times \mathbf{A}^{T}\mathbf{x}[n]\mathbf{x}[n]^{H}\mathbf{A}^{\ast }\mathbf{E}^{H}[n]\left (-j2\pi nT \right )\mathbf{U}^{H}\mathbf{e}_{i} \Big \}\Big ]\\
		&=2\mathrm{Re}\Big \{(\mathbf{B}^{H}\mathbf{Q}^{-1}\mathbf{B})_{ij} (\mathbf{U}^{\ast }\mathbf{A}^{H}\mathbf{R}_{x}^{\ast }\mathbf{A}\mathbf{U})_{ij}\\
		&\times(\mathbf{\Sigma }_{3})_{ij}\Big \}, \forall i,j\in \mathcal{M},
	\end{aligned}
 \end{equation}
where $(\cdot )_{ij}$ refers to the $i$th row and $j$th column element of the matrix. Thus we obtain $\mathbf{F}_{\mathcal{F}_{D} \mathcal{F}_{D}}=2\mathrm{Re}(\mathbf{F}_{44})$, with $\mathbf{F}_{44}$ specified in (\ref{eq:f}). Other terms of FIM can be calculated in the same way with the corresponding partial derivative.
Hence, we obtain the FIM in (\ref{fisher}).

\bibliographystyle{IEEEtran}  
\bibliography{reference}

\end{document}